# Approximate relativistic bound state solutions of the Tietz-Hua rotating oscillator for any $\kappa$ -state

Sameer M. Ikhdair[1*], Majid Hamzavi[2**]

[1]*Physics Department, Near East University, Nicosia, North Cyprus, Mersin 10, Turkey*

[2]*Department of Basic Sciences, Shahrood Branch, Islamic Azad University, Shahrood, Iran*

[*] *sikhdair@neu.edu.tr*

[**] Corresponding Author: *majid.hamzavi@gmail.com*

## Abstract

Approximate analytical solutions of the Dirac equation with Tietz-Hua (TH) potential are obtained for arbitrary spin-orbit quantum number $\kappa$ using the Pekeris approximation scheme to deal with the spin-orbit coupling terms $\kappa(\kappa \pm 1)r^{-2}$. In the presence of exact spin and pseudo-spin (pspin) symmetric limitation, the bound state energy eigenvalues and associated two-component wave functions of the Dirac particle moving in the field of attractive and repulsive TH potential are obtained using the parametric generalization of the Nikiforov-Uvarov (NU) method. The cases of the Morse potential, the generalized Morse potential and non-relativistic limits are studied.

## 1- Introduction

The spin and the pseudo-spin (pspin) symmetries of the Dirac Hamiltonian had been discovered many years ago, however, these symmetries have recently been recognized empirically in nuclear and hadronic spectroscopes [1]. Within the framework of Dirac equation, pspin symmetry used to feature deformed nuclei and superdeformation to establish an effective shell-model [2-4], whereas spin symmetry is relevant for mesons [5]. The spin symmetry occurs when the difference of the scalar $S(r)$ and vector $V(r)$ potentials are a constant, i.e., $\Delta(r) = C_s$ and the pspin symmetry occurs when the sum of the scalar and vector potentials are a constant, i.e., $\Sigma(r) = C_{ps}$ [6-7]. The pspin symmetry refers to a quasi-degeneracy of single nucleon doublets with non-relativistic quantum number $(n, l, j = l + 1/2)$ and $(n-1, l+2, j = l+3/2)$, where $n$, $l$ and $j$ are single nucleon radial, orbital and total angular quantum numbers, respectively [8-9]. The total angular momentum is $j = \tilde{l} + \tilde{s}$, where $\tilde{l} = l + 1$ is the pseudo-angular momentum and $\tilde{s}$ is the pspin angular momentum [10].

The Tietz-Hua (TH) oscillatory potential is one of the most suitable molecular potentials to describe the vibration energy spectra of diatomic molecules [11,12]. It is much more realistic than the generalized Morse potential in the description of molecular dynamics at moderate and high rotation-vibration quantum numbers [13-15]. Kunc and Gordillo-Vázquez derived analytical expressions for the rotation-vibration energy levels of diatomic molecules represented by the TH rotating oscillator potential using the Hamilton-Jacobi theory and the Bohr-Sommerfeld quantization rule [16]. The TH potential takes the following form:

$$V_{TH}(r) = D \left[ \frac{1 - e^{-b_h (r-r_e)}}{1 - c_h e^{-b_h (r-r_e)}} \right]^2, \quad b_h = \beta(1 - c_h), \qquad (1)$$

where the parameters $r$, $r_e$, $\beta$, $D$ and $c_h$ are the inter-nuclear distance, the molecular bond length, the Morse constant, the potential well depth and the potential constant, respectively [16]. In the limit when the potential constant $c_h$ approaches to zero, the TH potential turns to become the Morse potential [17]. In Figure 1, we draw the potential form in (1) for three different types of molecular potentials, namely, the TH, Morse and generalized Morse potentials. The following set of parameter values $r_e = 0.4\,fm$, $c_h = 0.1$, $D = 15.0\,fm^{-1}$ and $b_h = 0.8\,fm^{-1}$ are used.

Over the past years, the Nikiforov-Uvarov (NU) method [18] has shown to be a powerful tool in solving second-order differential equations. It was applied successfully to a large number of potential models [19-25]. This method has also been used to solve the spinless (spin-$0$) Schrödinger [26-30] and Klein-Gordon (KG) [31-35] equations and also relativistic spin-$1/2$ Dirac equation [36-40] with different potential models.

Since the relativistic solution is indispensable, we need to solve the Dirac equation with flexible parameters TH molecular potential model. However, the Dirac-Tietz-Hua problem can no longer be solved in a closed form due to the existence of spin-orbit coupling term $\kappa(\kappa\pm1)r^{-2}$ and it is necessary to resort to approximation methods. Therefore, we use Pekeris approximation scheme to deal with this term and solve approximately the Dirac equation with the Tietz-Hua potential for arbitrary spin-orbit quantum number $\kappa$. In the presence of exact spin and pseudo-spin symmetric limitation, we obtain the approximate relativistic bound state solutions including the energy eigenvalue equation and the corresponding unnormalized upper- and lower-spinor components of the wave functions using the concepts of parametric generalization of the NU method [41] since the relativistic corrections are not neglected. Furthermore, we consider a few special cases of interest like the Morse potential [42-44] the generalized Morse potential [41] and the non-relativistic limit of the present solution [45].

The structure of the paper is as follows. In Section 2, in the context of spin and pspin symmetry, we briefly introduce the Dirac equation with scalar and vector TH potentials for arbitrary spin-orbit quantum number $\kappa$. The Pekeris approximation scheme for the spin-orbit centrifugal and pseudo-centrifugal terms is presented in Appendix A. The parametric generalization of the NU method is displayed in appendix B. In the presence of the exact spin and pspin symmetry, the approximate energy eigenvalue equations and corresponding two-component wave functions of the Dirac-Tietz-Hua problem are obtained. In Section 3, we consider some particular cases of our solutions. Finally, our conclusion is given in Section 4.

## 2. Bound State Solutions

The Dirac equation for fermionic massive spin-$1/2$ particles moving in the field of an attractive scalar $S(r)$ and a repulsive vector $V(r)$ potential (in units $\hbar = c = 1$) is

$$\left[\vec{\alpha}\cdot\vec{p} + \beta\left(M + S(r)\right)\right]\psi(\vec{r}) = \left[E - V(r)\right]\psi(\vec{r}), \quad (2)$$

where $E$ is the relativistic energy of the system, $\vec{p} = -i\vec{\nabla}$ is the three-dimensional (3D) momentum operator and $M$ is the mass of the fermionic particle. $\vec{\alpha}$ and $\beta$ are the usual $4\times 4$ Dirac matrices [46]. One may closely follow the procedure described in Eqs. (17)–(19) of Ref. [47] to obtain

$$\left[\frac{d^2}{dr^2}-\frac{\kappa(\kappa+1)}{r^2}\right]F_{n\kappa}(r)+\frac{\frac{d\Delta(r)}{dr}}{M+E_{n\kappa}-\Delta(r)}\left(\frac{d}{dr}+\frac{\kappa}{r}\right)F_{n\kappa}(r)$$
$$=\left[(M+E_{n\kappa}-\Delta(r))(M-E_{n\kappa}+\Sigma(r))\right]F_{n\kappa}(r),\ r\in(0,\infty) \quad (3)$$

$$\left[\frac{d^2}{dr^2}-\frac{\kappa(\kappa-1)}{r^2}\right]G_{n\kappa}(r)+\frac{\frac{d\Sigma(r)}{dr}}{M-E_{n\kappa}+\Sigma(r)}\left(\frac{d}{dr}-\frac{\kappa}{r}\right)G_{n\kappa}(r)$$
$$=\left[(M+E_{n\kappa}-\Delta(r))(M-E_{n\kappa}+\Sigma(r))\right]G_{n\kappa}(r),\ r\in(0,\infty) \quad (4)$$

where $\kappa(\kappa-1)=\tilde{l}(\tilde{l}+1)$ and $\kappa(\kappa+1)=l(l+1)$. The orbit-spin quantum number $\kappa$ is related to the orbital quantum numbers $l$ and $\tilde{l}$ for spin symmetry and pspin symmetric models, respectively, as

$$\kappa=\begin{cases} -(l+1)=-(j+\frac{1}{2}) & (s_{1/2},p_{3/2},etc.)\ j=l+\frac{1}{2},\ \text{aligned spin}\ (\kappa\langle 0), \\ +l=+(j+\frac{1}{2}) & (p_{1/2},d_{3/2},etc.)\ j=l-\frac{1}{2},\ \text{unaligned spin}\ (\kappa\rangle 0). \end{cases}$$

Further, $\kappa$ in the quasi-degenerate doublet structure can be expressed in terms of $\tilde{s}=1/2$ and $\tilde{l}$, the pspin and pseudo-orbital angular momentum, respectively, as

$$\kappa=\begin{cases} -\tilde{l}=-(j+\frac{1}{2}) & (s_{1/2},p_{3/2},etc.)\ j=\tilde{l}-\frac{1}{2},\ \text{aligned pseudospin}\ (\kappa\langle 0), \\ +(\tilde{l}+1)=+(j+\frac{1}{2}) & (d_{3/2},f_{5/2},etc.)\ j=\tilde{l}+\frac{1}{2},\ \text{unaligned spin}\ (\kappa\rangle 0), \end{cases}$$

where $\kappa=\pm 1,\pm 2,\ldots$ For example, the states $(1s_{1/2},0d_{3/2})$ and $(1p_{3/2},0f_{5/2})$ can be considered as pseudospin doublets.

## 2.1. Spin Symmetric Limit

In the spin symmetric limitation, $\frac{d\Delta(r)}{dr}=0$ or $\Delta(r)=C_s=\text{constant}$ [48,49], then Eq. (3) with $\Sigma(r)=V_{\text{TH}}(r)$, becomes

$$\left[\frac{d^2}{dr^2} - \frac{\kappa(\kappa+1)}{r^2} - \gamma D \left(\frac{1-e^{-b_h(r-r_e)}}{1-c_h e^{-b_h(r-r_e)}}\right)^2 - \beta^2\right] F_{n\kappa}(r) = 0, \tag{5a}$$

$$\gamma = M + E_{n\kappa} - C_s \text{ and } \beta^2 = (M - E_{n\kappa})(M + E_{n\kappa} - C_s). \tag{5b}$$

where $\kappa = l$ and $\kappa = -l-1$ for $\kappa < 0$ and $\kappa > 0$, respectively. The Schrödinger-like equation (5a) that results from the Dirac equation is a second order differential equation containing a spin-orbit centrifugal term $\kappa(\kappa+1)r^{-2}$ which is singular at $r=0,$ and needs to be treated very carefully while performing the approximation. The widely known Pekeris approximation (see Appendix A) in which the spin-orbit centrifugal coupling term $\kappa(\kappa+1)r^{-2}$ is expanded in terms of singular functions of $e^{-r/a}$ compatible with the solvability of the problem for $r \ll a.$ Because of this singularity, the validity of such approximation is limited only to very few of the lowest energy states.

Equation (5a) has an exact rigorous solution only for the states with $\kappa = -1$ because of the existence of the centrifugal term $\kappa(\kappa+1)/r^2$. However, when this term is taken into account, the corresponding radial Dirac equation can no longer be solved in a closed form and it is necessary to resort to approximate methods. Over the last few decades several schemes have been used to calculate the energy spectrum. The main idea of these schemes relies on using different approximations of the spin-orbit centrifugal coupling term $\kappa(\kappa+1)/r^2$. So we need to perform a new approximation for the spin-orbit term as a function of the TH potential parameters in Appendix A. This approximation in the limit when $c_h = 0$ reduces to the approximation used before for the Morse potential case [42]. Thus, employing such an approximation scheme and making change of variables $x = (r-r_e)/r_e \in (-1, \infty)$, we can then write Eq. (5) as:

$$\left[\frac{d^2}{dx^2} - \kappa(\kappa+1)\left(D_0 + D_1 \frac{e^{-\alpha x}}{1-c_h e^{-\alpha x}} + D_2 \frac{e^{-2\alpha x}}{(1-c_h e^{-\alpha x})^2}\right)\right.$$

$$\left. - \gamma D r_e^2 \left(\frac{1-e^{-\alpha x}}{1-c_h e^{-\alpha x}}\right)^2 - \beta^2 r_e^2 \right] F_{n\kappa}(x) = 0, \tag{6}$$

where the explicit forms of the constants $D_i$ $(i=1,2 \text{ and } 3)$ are defined in (A3) and expressed in terms of the potential parameters $(c_h, b_h, r_e)$. Followed by setting a new variable $s(r) = e^{-\alpha x} \in (e^{b_h r_e}, 0)$, this allows us to decompose the spin-symmetric Dirac equation (6) into the Schrödinger-type equation satisfying the upper-spinor component $F_{n,\kappa}(s)$,

$$\left\{\frac{d^2}{ds^2}+\frac{1-c_h s}{s(1-c_h s)}\frac{d}{ds}-\frac{1}{s^2(1-c_h s)^2}\left[\frac{a_1\kappa(\kappa+1)}{\alpha^2}+\frac{a_2 r_e^2}{\alpha^2}\right]\right\}F_{n,\kappa}(s)=0,$$

$$a_1=\left(D_0(1-c_h s)^2+D_1 s(1-c_h s)+D_2\right),\ a_2=\left(\gamma D(1-s)^2+\beta^2(1-c_h s)^2\right), \qquad (7)$$

where $F_{n\kappa}(x)\equiv F_{n,\kappa}(s)$ has been used. If the above equation is compared with (B2), we can obtain the specific values for constants $c_i$ ($i=1,2,3$) along with $\xi_j$ ($j=1,2,3$) as

$c_1=1,\ c_2=c_h,\ c_3=c_h,$

$$\xi_1=\frac{1}{\alpha^2}\left[\kappa(\kappa+1)\left(D_0 c_h^2-D_1 c_h+D_2\right)+\gamma D r_e^2+\beta^2 r_e^2 c_h^2\right],$$

$$\xi_2=\frac{1}{\alpha^2}\left[-\kappa(\kappa+1)\left(-2D_0 c_h+D_1\right)+2\gamma D r_e^2+2\beta^2 r_e^2 c_h\right],$$

$$\xi_3=\frac{1}{\alpha^2}\left[-\kappa(\kappa+1)D_0+\gamma D r_e^2+2\beta^2 r_e^2\right].$$

In order to obtain the bound state solutions of Eq. (7), it is necessary to calculate the remaining parametric constants, that is, $c_i$ ($i=4,5,...,13$) by means of the relation (B5). Their specific values are displayed in table 1 for the relativistic TH potential model. Further, using these constants along with (B10), we can readily obtain the energy eigenvalue equation for the Dirac-TH problem as

$$(2n+1)\left(\sqrt{\frac{c_h^2}{4}+\frac{1}{\alpha^2}\left[\kappa(\kappa+1)D_2+\gamma D r_e^2(c_h-1)^2\right]}+c_h\sqrt{\frac{1}{\alpha^2}\left[-\kappa(\kappa+1)D_0+\gamma D r_e^2+2\beta^2 r_e^2\right]}\right)$$

$$+2\sqrt{\left(\frac{c_h^2}{4}+\frac{1}{\alpha^2}\left[\kappa(\kappa+1)D_2+\gamma D r_e^2(c_h-1)^2\right]\right)\left(\frac{1}{\alpha^2}\left[-\kappa(\kappa+1)D_0+\gamma D r_e^2+2\beta^2 r_e^2\right]\right)}$$

$$+\frac{\kappa(\kappa+1)}{\alpha^2}D_1+\frac{2\gamma D r_e^2}{\alpha^2}(c_h-1)+\frac{c_h}{2}+c_h n(n-1)=0. \qquad (8)$$

Recalling $\gamma=M+E_{n\kappa}-C_s$ and $\beta^2=(M-E_{n\kappa})(M+E_{n\kappa}-C_s)$ from Eq. (5b), one can obtain the implicit dependence of the above energy equation on the energy $E_{n\kappa}$. In order to establish the upper-spinor component of the wave functions $F_{n,\kappa}(r)$, namely,

Eq. (5a), the relations (B11)-(B14) are used. Firstly, we find the first part of the wave function as

$$\phi(s) = s^{\sqrt{\frac{1}{\alpha^2}\left[-\kappa(\kappa+1)D_0+\gamma Dr_e^2+2\beta^2 r_e^2\right]}} (1-c_h s)^{\frac{1}{2}+\frac{1}{c_h}\sqrt{\frac{c_h^2}{4}+\frac{1}{\alpha^2}\left[\kappa(\kappa+1)D_2+\gamma Dr_e^2(c_h-1)^2\right]}}. \tag{9}$$

Secondly, we calculate the weight function as

$$\rho(s) = s^{2\sqrt{\frac{1}{\alpha^2}\left[-\kappa(\kappa+1)D_0+\gamma Dr_e^2+2\beta^2 r_e^2\right]}} (1-c_h s)^{\frac{2}{c_h}\sqrt{\frac{c_h^2}{4}+\frac{1}{\alpha^2}\left[\kappa(\kappa+1)D_2+\gamma Dr_e^2(c_h-1)^2\right]}}, \tag{10}$$

which gives the second part of the wave function as

$$y_n(s) = P_n^{\left(2\sqrt{\frac{1}{\alpha^2}\left[-\kappa(\kappa+1)D_0+\gamma Dr_e^2+2\beta^2 r_e^2\right]},\; \frac{2}{c_h}\sqrt{\frac{c_h^2}{4}+\frac{1}{\alpha^2}\left[\kappa(\kappa+1)D_2+\gamma Dr_e^2(c_h-1)^2\right]}\right)}(1-2c_h s), \tag{11}$$

where $P_n^{(a,b)}(1-2c_h s)$ are the orthogonal Jacobi polynomials. Finally the upper spinor component for arbitrary $\kappa$ can be found through the relation (B14)

$$F_{n\kappa}(s) = N_{n\kappa} s^{\sqrt{\frac{1}{\alpha^2}\left[-\kappa(\kappa+1)D_0+\gamma Dr_e^2+2\beta^2 r_e^2\right]}} (1-c_h s)^{\frac{1}{2}+\frac{1}{c_h}\sqrt{\frac{c_h^2}{4}+\frac{1}{\alpha^2}\left[\kappa(\kappa+1)D_2+\gamma Dr_e^2(c_h-1)^2\right]}}$$

$$\times P_n^{\left(2\sqrt{\frac{1}{\alpha^2}\left[-\kappa(\kappa+1)D_0+\gamma Dr_e^2+2\beta^2 r_e^2\right]},\; \frac{2}{c_h}\sqrt{\frac{c_h^2}{4}+\frac{1}{\alpha^2}\left[\kappa(\kappa+1)D_2+\gamma Dr_e^2(c_h-1)^2\right]}\right)}(1-2c_h s), \tag{12}$$

where $N_{n\kappa}$ is the normalization constant. On the other hand, the lower-spinor component of the wave function can be calculated by using

$$G_{n\kappa}(r) = \frac{1}{M+E_{n\kappa}-C_s}\left(\frac{d}{dr}+\frac{\kappa}{r}\right)F_{n\kappa}(r), \tag{13}$$

where $E \neq -M + C_s$ and in the presence of the exact spin symmetry ($C_s = 0$), only positive energy states do exist.

## 2.2. Pseudospin Symmetric Limit

Ginocchio showed that there is pspin symmetry in case when the relationship between the vector potential and the scalar potential is given by $V(r) = -S(r)$ [7]. Further, Meng et al. showed that if $\frac{d[V(r)+S(r)]}{dr} = \frac{d\Sigma(r)}{dr} = 0$, then $\Sigma(r) = C_{ps} = $ constant, for which the pspin symmetry is exact in the Dirac equation [48-49]. Thus, choosing the $\Delta(r)$ as the TH potential, Eq. (4) under this symmetry becomes

$$\left[\frac{d^2}{dr^2} - \frac{\kappa(\kappa-1)}{r^2} - \tilde{\gamma}D\left(\frac{1-e^{-b_h(r-r_e)}}{1-c_h e^{-b_h(r-r_e)}}\right)^2 - \tilde{\beta}^2\right]G_{n\kappa}(r) = 0, \tag{14a}$$

$$\tilde{\gamma} = E_{n\kappa} - M - C_{ps} \text{ and } \tilde{\beta}^2 = (M + E_{n\kappa})(M - E_{n\kappa} + C_{ps}),. \tag{14b}$$

where $\kappa = -\tilde{l}$ and $\kappa = \tilde{l}+1$ for $\kappa < 0$ and $\kappa > 0$, respectively. Employing the new approximation derived for the spin-orbit pseudo-centrifugal term, $\kappa(\kappa-1)/r^2$ in Appendix A, the pspin Dirac equation (14a) can be written as

$$\left[\frac{d^2}{dx^2} - \kappa(\kappa-1)\left(D_0 + D_1\frac{e^{-\alpha x}}{1-c_h e^{-\alpha x}} + D_2\frac{e^{-2\alpha x}}{(1-c_h e^{-\alpha x})^2}\right)\right.$$

$$\left. -\tilde{\gamma}Dr_e^2\left(\frac{1-e^{-\alpha x}}{1-c_h e^{-\alpha x}}\right)^2 - \tilde{\beta}^2 r_e^2\right]G_{n\kappa}(r) = 0 \tag{15}$$

To avoid repetition, the negative energy solution of Eq. (15), in the exact pspin symmetric case: $V(r) = -S(r)$, can be readily obtained directly via the spin symmetric solution throughout the following parametric mappings:

$$F_{n\kappa}(r) \leftrightarrow G_{n\kappa}(r), \kappa \to \kappa-1, V(r) \to -V(r) (i.e., D \to -D), E_{n\kappa} \to -E_{n\kappa}, C_s \to -C_{ps}. \tag{16}$$

Following the previous procedure, one can obtain the energy equation for the Dirac hole energy states in the presence of the pspin symmetric case as

$$(2n+1)\left(\sqrt{\frac{c_h^2}{4}+\frac{1}{\alpha^2}\left[\kappa(\kappa-1)D_2+\tilde{\gamma}Dr_e^2(c_h-1)^2\right]}+c_h\sqrt{\frac{1}{\alpha^2}\left[-\kappa(\kappa-1)D_0+\tilde{\gamma}Dr_e^2+2\tilde{\beta}^2 r_e^2\right]}\right)$$

$$+2\sqrt{\left(\frac{c_h^2}{4}+\frac{1}{\alpha^2}\left[\kappa(\kappa-1)D_2+\tilde{\gamma}Dr_e^2(c_h-1)^2\right]\right)\left(\frac{1}{\alpha^2}\left[-\kappa(\kappa-1)D_0+\tilde{\gamma}Dr_e^2+2\tilde{\beta}^2 r_e^2\right]\right)}$$

$$+\frac{\kappa(\kappa-1)}{\alpha^2}D_1+\frac{2\tilde{\gamma}Dr_e^2}{\alpha^2}(c_h-1)+\frac{c_h}{2}+c_h n(n-1) = 0. \tag{17}$$

It should be pointed out that the solutions of energy equation (17) are not for valence states but for Dirac hole states.

Furthermore, we calculate the lower-spinor component of the Dirac hole wave functions to be

$$G_{n\kappa}(s) = \tilde{N}_{n,\kappa} s^{\sqrt{\frac{1}{\alpha^2}\left[-\kappa(\kappa-1)D_0+\tilde{\gamma}Dr_e^2+2\tilde{\beta}^2 r_e^2\right]}} (1-c_h s)^{\frac{1}{2}+\frac{1}{c_h}\sqrt{\frac{c_h^2}{4}+\frac{1}{\alpha^2}\left[\kappa(\kappa-1)D_2+\tilde{\gamma}Dr_e^2(c_h-1)^2\right]}}$$

$$\times P_n^{\left(2\sqrt{\frac{1}{\alpha^2}\left[-\kappa(\kappa-1)D_0+\tilde{\gamma}Dr_e^2+2\tilde{\beta}^2 r_e^2\right]}, \frac{2}{c_h}\sqrt{\frac{c_h^2}{4}+\frac{1}{\alpha^2}\left[\kappa(\kappa-1)D_2+\tilde{\gamma}Dr_e^2(c_h-1)^2\right]}\right)}(1-2c_h s), \tag{18}$$

where $\tilde{N}_{n\kappa}$ is the normalization constant. Again, recalling $\tilde{\gamma} = E_{n\kappa} - M - C_{ps}$ and $\tilde{\beta}^2 = (M + E_{n\kappa})(M - E_{n\kappa} + C_{ps})$ from Eq. (14b), one can obtain the implicit dependence on the energy $E_{n\kappa}$. The upper-spinor component of the Dirac hole wave function can be calculated by

$$F_{n\kappa}(r) = \frac{1}{M - E_{n\kappa} + C_{ps}} \left( \frac{d}{dr} - \frac{\kappa}{r} \right) G_{n\kappa}(r), \tag{19}$$

where $E \neq M + C_{ps}$ and in the presence of the exact pspin symmetry ($C_{ps} = 0$), only negative energy states do exist.

## 3. A Few Special Cases

In this section we consider some special cases of interest from the TH potential as follows: First, when we change the potential parameters as $b_h = \alpha$ and $c_h = e^{-\alpha r_e}$, the potential reduces into the generalized Morse potential (GMP) proposed by Deng and Fan [50], i.e.;

$$V_{GM}(r) = D \left( 1 - \frac{b}{e^{\alpha r} - 1} \right)^2, \qquad b = e^{\alpha r_e} - 1. \tag{20}$$

Very recently, one of us [41] has studied the approximate spin (pseudospin) symmetry limitation of the bound state solutions of the Dirac equation with this potential for any arbitrary spin-orbit $\kappa$ using the parametric generalization of the NU method including a new improved approximation scheme to deal with spin-orbit barrier term.

Second, when $c_h = 0$, the TH potential reduces to the Morse potential (Version I), i.e.

$$\lim_{c_h \to 0} V_M(r) = V_M^{(I)}(r) = D_e (e^{-2b_h(r-r_e)} - 2e^{-b_h(r-r_e)}) + D_e. \tag{21}$$

Thus, we can obtain the two energy equations for the Dirac-Morse problem

$$\frac{2}{\alpha^2} \sqrt{\left[\kappa(\kappa+1)D_2 + \gamma D_e r_e^2\right]\left[-\kappa(\kappa+1)D_0 + \gamma D_e r_e^2 + 2\beta^2 r_e^2\right]}$$

$$+ \frac{(2n+1)}{\alpha} \sqrt{\kappa(\kappa+1)D_2 + \gamma D_e r_e^2} + \frac{1}{\alpha^2} \left[\kappa(\kappa+1)D_1 - 2\gamma D_e r_e^2\right] = 0, \tag{22}$$

and

$$\frac{2}{\alpha^2} \sqrt{\left[\kappa(\kappa-1)D_2 + \tilde{\gamma} D_e r_e^2\right]\left[-\kappa(\kappa-1)D_0 + \tilde{\gamma} D_e r_e^2 + 2\tilde{\beta}^2 r_e^2\right]}$$

$$+\frac{(2n+1)}{\alpha}\sqrt{\kappa(\kappa-1)D_2 + \tilde{\gamma}D_e r_e^2} + \frac{1}{\alpha^2}\left[\kappa(\kappa-1)D_1 - 2\tilde{\gamma}D_e r_e^2\right] = 0, \qquad (23)$$

for spin and pspin symmetric limitation, respectively. The constants $D_i$ ($i=1,2$ and $3$) when $c_h \to 0$ are also found in (A3) and expressed in terms of potential parameters $(b_h, r_e)$.

Also, when we neglect the last term in Eq. (21), we get the second version of the Morse potential (version II):

$$V_M^{(II)} = D_e(e^{-2b_h(r-r_e)} - 2e^{-b_h(r-r_e)}). \qquad (24)$$

Recently, Berkdemir [42-43] and Aydoğdu et al [44] have studied the above potential in the context of the relativistic theory.

To show the procedure of determining the energy eigenvalues from Eqs. (8) and (17), we take a set of physical parameter values, $r_e = 2.40873\,fm$, $b_h = 0.988879\,fm^{-1}$, $D = 5.0\,fm^{-1}$, $M = 10.0\,fm^{-1}$ and $C_s = 10.0$ [44]. At first, we test the accuracy of this potential model by comparing our numerical results with the two version of the Morse potential for various quantum numbers $n$ and $\kappa$. Hence, we display the approximated energy levels in tables 2 and 3. For example, in the presence of spin symmetry, table 2 presents the energy spectrum of the TH potential as well. Obviously, the pairs $(np_{1/2}, np_{3/2})$, $(nd_{3/2}, nd_{5/2})$, $(nf_{5/2}, nf_{7/2})$, $(ng_{7/2}, ng_{9/2})$, and so on are degenerate states. Thus, each pair is considered as spin doublet and has positive energy [41]. Also in table 3, we give the numerical results for the Dirac hole energy states for the pspin symmetric case. Here, we take the following set of parameter values, $r_e = 2.40873\,fm$, $b_h = 0.988879\,fm^{-1}$, $D = 5.0\,fm^{-1}$, $M = 10.0\,fm^{-1}$ and $C_{ps} = -10.0$ [44]. We observe the degeneracy in the following doublets $(1s_{1/2}, 0d_{3/2})$, $(1p_{3/2}, 0f_{5/2})$, $(1d_{5/2}, 0g_{7/2})$, $(1f_{7/2}, 0h_{9/2})$, and so on. Thus, each pair is considered as pspin doublet and has negative Dirac hole energy states [41].

Third, in our applications to the diatomic molecules, we use the potential parameters of $H_2$ and $I_2$ which are obtained from Ref. [16] and also displayed in table 4. Some numerical values for the energy levels of these two diatomic molecules are presented in tables 5 and 6 for spin and pspin symmetries, respectively. Note that we used $\hbar c = 1973.29\,eV\mathring{A}$ [21,30] with the choice of $1\,amu = 931.494028\,MeV/c^2$ [45].

Fourth, we study the energy eigenvalue equation (8) and upper-spinor component of wave function (12) of the Dirac-TH problem under the nonrelativistic limits $E_{n\kappa} - M \to E_{nl}$ and $M + E_{n\kappa} \to 2\mu$. Thus, we obtain the energy equation of the Schrödinger equation with any arbitrary orbital state for the TH potential as

$$c_h\left(n+\frac{1}{2}\right)^2 + \frac{c_h}{4} + 2\left(n+\frac{1}{2}\right)\left(\sqrt{\xi_1 - c_h\xi_2 + c_h^2\xi_3 + \frac{c_h^2}{4}} + c_h\sqrt{\frac{r_e^2}{\alpha^2}(d-\varepsilon) + \frac{l(l+1)}{\alpha^2}D_0}\right)$$

$$-\frac{2r_e^2}{\alpha^2}(-\varepsilon c_h + d) + \frac{l(l+1)}{\alpha^2}(-2D_0 c_h + D_1) - 2c_h\left(\frac{r_e^2}{\alpha^2}(d-\varepsilon) + \frac{l(l+1)}{\alpha^2}D_0\right)$$

$$+2\sqrt{\left(\frac{r_e^2}{\alpha^2}(d-\varepsilon) + \frac{l(l+1)}{\alpha^2}D_0\right)\left(\xi_1 - c_h\xi_2 + c_h^2\xi_3 + \frac{c_h^2}{4}\right)} = 0, \tag{25}$$

and the radial wave function as

$$R_{n,l}(s) = s^{\sqrt{\frac{r_e^2}{\alpha^2}(d-\varepsilon) + \frac{l(l+1)}{\alpha^2}D_0}}(1-s)^{\frac{1}{2} + \frac{1}{c_h}\sqrt{\xi_1 - c_h\xi_2 + c_h^2\xi_3 + \frac{c_h^2}{4}}}$$

$$\times P_n^{(2\sqrt{\frac{r_e^2}{\alpha^2}(d-\varepsilon) + \frac{l(l+1)}{\alpha^2}D_0}, \frac{2}{c_h}\sqrt{\xi_1 - c_h\xi_2 + c_h^2\xi_3 + \frac{c_h^2}{4}})}(1 - 2c_h s), \tag{26}$$

where $\varepsilon = 2\mu E_{nl}/\hbar^2$, $d = 2\mu D/\hbar^2$ and

$$\xi_1 = \frac{r_e^2}{\alpha^2}(d - \varepsilon c_h^2) + \frac{l(l+1)}{\alpha^2}(D_0 c_h^2 - D_1 c_h + D_2) \tag{27a}$$

$$\xi_2 = \frac{2r_e^2}{\alpha^2}(-\varepsilon c_h + d) - \frac{l(l+1)}{\alpha^2}(-2D_0 c_h + D_1) \tag{27b}$$

$$\xi_3 = \frac{r_e^2}{\alpha^2}(d - \varepsilon) + \frac{l(l+1)}{\alpha^2}D_0 \tag{27c}$$

Very recently, the results for this case are presented by Hamzavi et al. [45].

Finally, we plot the relativistic energy eigenvalues of the TH potential under spin and pspin symmetry limitations for the different ground degenerate spin doublet levels $(0p_{1/2}, 0p_{3/2})$ with $\kappa = 1, -2$ and $(0d_{3/2}, 0d_{5/2})$ with $\kappa = 2, -3$ and also first excited degenerate spin doublet levels $(1f_{5/2}, 1f_{7/2})$ with $\kappa = 3, -4$ and $(1g_{7/2}, 1g_{9/2})$ with $\kappa = 4, -5$ in Figures 2 to 7. Figure 2 shows the plot the energy eigenvalues of spin symmetry limit versus the potential parameter $b_h$. It is noted that when $b_h$ is increasing, the energy is slightly decreasing. Figure 3 shows the variation of the energy as a function of $c_h$. When the parameter $c_h$ increases, the energy increases, too for $c_h > 0.1$ and $c_h < -0.1$. However, it is found that the change is very slow for the interval $-0.1 < c_h < 0.1$. Also, we plot the energy as a function of $r_e$ in Fig. 4. It is noted that when $r_e$ increasing, the energy is sharply decreasing. In Figs. 5 to 7, we plot the Dirac hole energy states of the pspin symmetry limit as functions of parameters $b_h$, $c_h$ and $r_e$, respectively. The variation of energy can also be seen slightly increasing, decreasing (in $c_h > 0.1$ and $c_h < -0.1$) and sharply increasing, respectively. The behavior in Figs. 5 to 7 is found to be opposite to the spin symmetric case.

## 4. Conclusion

In this work, we have investigated the bound state solutions of the Dirac equation with Tietz-Hua potential for any spin-orbit quantum number $\kappa$. By making a Pekeris approximation to deal with the spin-orbit centrifugal (pseudo-centrifugal) coupling term, we obtain the energy eigenvalue equation and the unnormalized two components of the radial wave function expressed in terms of the Jacobi polynomials. The spin and pspin symmetry cases are studied for any $\kappa$ wave state. In this regard, numerical values for the Dirac valence (hole) energy states are obtained for the spin (pspin) case. Furthermore, our analytical solutions can be reduced to the relativistic generalized Morse solution by simply making a proper change of parameters $b_h = \alpha$ and $c_h = e^{-\alpha r_e}$ and to the Morse potential by inserting $c_h = 0$. The relativistic solution can be reduced into the Schrödinger solution in the nonrelativistic limit. Our numerical results are compared with the existing energy spectra for the particular case when $c_h = 0$ (the Morse case).

## Acknowledgments

We would like to thank the kind referee for positive suggestions which have greatly improved the present work.

## References

[1] J. N. Ginocchio, Phys. Rep. **414** (4-5) (2005) 165.
[2] A. Bohr, I. Hamamoto and B. R. Mottelson, Phys. Scr. **26** (1982) 267.
[3] J. Dudek, W. Nazarewicz, Z. Szymanski and G. A. Leander, Phys. Rev. Lett. **59** (1987) 1405.

[4] D. Troltenier, C. Bahri and J. P. Draayer, Nucl. Phys. A **586** (1995) 53.

[5] P. R. Page, T. Goldman and J. N. Ginocchio, Phys. Rev. Lett. **86** (2001) 204.

[6] J. N. Ginocchio, A. Leviatan, J. Meng, and S. G. Zhou, Phys. Rev. C **69** (2004) 034303.

[7] J. N. Ginocchio, Phys. Rev. Lett. **78** (3) (1997) 436.

[8] K. T. Hecht and A. Adler, Nucl. Phys. A **137** (1969) 129.

[9] A. Arima, M. Harvey and K. Shimizu, Phys. Lett. B **30** (1969) 517.

[10] S. M. Ikhdair and R. Sever, Appl. Math. Com. **216** (2010) 911.

[11]. T. Tietz, J. Chem. Phys. **38** (1963) 3036.

[12]. W. Hua, Phys. Rev. A **24** (1990) 2524.

[13]. G. A. Natanson, Phys. Rev. A **44** (1991) 3377.

[14]. E. Levin, H. Partridge and J. R. Stallcop, J. Thermo Phys. Heat Transfer **4** (1990) 469.

[15]. R. T. Pack, J. Chem. Phys. **57** (1972) 4612.

[16]. J. A. Kunc and F. J. Gordillo-Vázquez, J. Phys. Chem. A **101** (1997) 1595.

[17]. P. M. Morse, Phys. Rev. **34** (1929) 57.

[18]. A. F. Nikiforov and V. B. Uvarov: Special Functions of Mathematical Physic, Birkhausr, Berlin, 1988.

[19] F. Yaşuk, C. Berkdemir and A. Berkdemir, J. Phys. A: Math. Gen. **38** (2005) 6579.

[20] M. R. Setare and S. Haidari, Phys. Scr. **81** (2010) 015201.

[21] S. M. Ikhdair and R. Sever, J. Math. Chem. **42** (2006) 461.

[22] M. G. Miranda, G. H. Sun and S. H. Dong, Int. J. Mod. Phys. E **19** (2010) 123.

[23] M. Hamzavi and A. A. Rajabi, Z. Naturforsch. **66a** (2011) 533.

[24] M. Hamzavi, H. Hassanabadi and A. A. Rajabi, Int. J. Mod. Phys. E **19** (2010) 2189.

[25] M. Hamzavi, H. Hassanabadi and A. A. Rajabi, Int. J. Theor. Phys. **50** (2011) 454.

[26] S. M. Ikhdair and R. Sever, Int. J. Mod. Phys. A **25** (2010) 3941.

[27] S. M. Ikhdair and J. Abu-Hasna, Phys. Scr. **83** (2011) 025002.

[28] Y. F. Cheng and T. Q. Dai Phys. Scr. **75** (2007) 274.

[29] S. M. Ikhdair and R. Sever, Ann. Phys. (Berlin) **16** (2007) 218.

[30] M. Aktaş and R. Sever, J. Mol. Struct.: Theochem **710** (2004) 223; S. M. Ikhdair and R. Sever, J. Mol. Struct.: Theochem **806** (2007) 155

[31] M. Şimşek and H. Eğrifes, J. Phys. A: Math. Gen. **37** (2004) 4379.

[32] S. M. Ikhdair and R. Sever, Int. J. Mod. Phys. C **19** (2008) 1425.

[33] H. Eğrifes and R. Sever, Int. J. Theor. Phys. **46** (2007) 935.

[34] S. M. Ikhdair and R. Sever, Phys. Scr. **79** (2009) 035002; S. M. Ikhdair, J. Quantum Infor. Science **1** (2011) 73.

[35] N. Saad, Phys. Scr. **76** (2007) 623.

[36] M. Hamzavi, A. A. Rajabi and H. Hassanabadi, Few-Body Syst. **48** (2010) 171; ibid, Few-Body Syst. **48** (2012) 171.

[37] M. Hamzavi, A. A. Rajabi and H. Hassanabadi, Phys. Lett. A **374** (2010) 4303.

[38] M. Hamzavi, A. A. Rajabi and H. Hassanabadi, Int. J. Mod. Phys. A **26** (2011) 1363.

[39] S. M. Ikhdair and R. Sever, Cent. Eur. J. Phys. **8** (2010) 665; S. M. Ikhdair and R. Sever, J. Math. Phys. **52** (2011) 122108.

[40] S. M. Ikhdair and R. Sever, J. Phys. A: Math. and Theor. **44** (2011) 355301.

[41] S. M. Ikhdair, J. Math. Phys. **52** (2011) 052303.

[42] C. Berkdemir, Nucl. Phys. A **770** (2006) 32.

[43] C. Berkdemir, Nucl. Phys. A **821** (2009) 262.

[44] O. Aydoğdu and R. Sever, Phys. Lett. B **703** (2011) 379.

[45] M. Hamzavi, A. A. Rajabi and H. Hassanabadi, submitted to Mol. Phys.

[46] W. Greiner, Relativistic Quantum Mechanics, Wave Equations, Third Edition, Springer, 2000.

[47] S. M. Ikhdair, J. Math. Phys. **51** (2010) 023525.

[48] J. Meng, K. Sugawara-Tanabe, S. Yamaji and A. Arima, Phys. Rev. C **59** (1999) 154; J. Meng, K. Sugawara-Tanabe, S. Yamaji, P. Ring and A. Arima, Phys. Rev. C **58** (1998) R628.

[49] S. G. Zhou, J. Meng, and P. Ring, Phys. Rev. Lett. 91 (2003) 262501. X. T. He, S. G. Zhou, J. Meng, E. G. Zhao, and W. Scheid, Euro. Phys. J. A 28 (2006) 265.

[50] Z. H. Deng and Y. P. Fan, Shandong Univ. J. **7** (1957) 162.

[51] S. M. Ikhdair, Int. J. Mod. Phys. C **20** (10) (2009) 1563.

**Appendix A: Pekeris Approximation to the Centrifugal Term**

Here we make a new approximation in order to deal with the spin-orbit centrifugal and pseudo centrifugal coupling terms given in Eqs. (5a) and (14a), respectively. The centrifugal (pseudo centrifugal) term is expanded around $r = r_e$ in a series of powers of $x = (r - r_e)/r_e \in (-1, \infty)$, as

$$V_{so}(r) = \frac{\eta_\kappa}{r^2} = \frac{\eta_\kappa}{r_e^2 (1+x)^2} = \frac{\eta_\kappa}{r_e^2}(1 - 2x + 3x^2 - 4x^3 + \ldots), \quad x \ll 1, \quad (A1)$$

where $\eta_\kappa = \kappa(\kappa \pm 1)$. It is sufficient to keep expansion terms only up to the second order. The above spin-orbit potential (A1) can be replaced by the form formally homogeneous to the original TH potential to keep the factorizability of the corresponding Schrödinger-like equation. Taking the centrifugal (pseudo centrifugal) term as

$$\tilde{V}_{so}(r) = \frac{\eta_\kappa}{r_e^2}\left[D_0 + D_1 \frac{e^{-\alpha x}}{1 - c_h e^{-\alpha x}} + D_2 \frac{e^{-2\alpha x}}{(1 - c_h e^{-\alpha x})^2}\right], \quad x \ll 1/\alpha, \quad \alpha x \ll 1 \quad (A2)$$

where $\alpha = b_h r_e$ and $D_i$ are the parameter of coefficients ($i = 0, 1, 2$). After making a Taylor expansion to (A2) up to the second order term, $x^2$, and then comparing equal powers with (A1), we can readily determine $D_i$ parameters as a function of the specific potential parameters $b_h$, $c_h$ and $r_e$ as follows:

$$D_0 = 1 - \frac{1}{\alpha}(1 - c_h)(3 + c_h) + \frac{3}{\alpha^2}(1 - c_h)^2, \quad \lim_{c_h \to 0} D_0 = 1 - \frac{3}{\alpha} + \frac{3}{\alpha^2}, \quad (A3a)$$

$$D_1 = \frac{2}{\alpha}(1 - c_h)^2(2 + c_h) - \frac{6}{\alpha^2}(1 - c_h)^3, \quad \lim_{c_h \to 0} D_1 = \frac{4}{\alpha} - \frac{6}{\alpha^2}, \quad (A3b)$$

$$D_2 = -\frac{1}{\alpha}(1 - c_h)^3(1 + c_h) + \frac{3}{\alpha^2}(1 - c_h)^4, \quad \lim_{c_h \to 0} D_2 = -\frac{1}{\alpha} + \frac{3}{\alpha^2}. \quad (A3c)$$

**Appendix B: Parametric Generalization of the NU method**

The NU method is used to solve second order differential equations with an appropriate coordinate transformation $s = s(r)$ [18]

$$\psi_n''(s) + \frac{\tilde{\tau}(s)}{\sigma(s)}\psi_n'(s) + \frac{\tilde{\sigma}(s)}{\sigma^2(s)}\psi_n(s) = 0, \quad (B1)$$

where $\sigma(s)$ and $\tilde{\sigma}(s)$ are polynomials, at most of second degree, and $\tilde{\tau}(s)$ is a first-degree polynomial. To make the application of the NU method simpler and direct without need to check the validity of solution. We present a shortcut for the method. So, at first we write the general form of the Schrödinger-like equation (B1) in a more general form applicable to any potential as follows [51]

$$\psi_n''(s) + \left(\frac{c_1 - c_2 s}{s(1 - c_3 s)}\right)\psi_n'(s) + \left(\frac{-\xi_1 s^2 + \xi_2 s - \xi_3}{s^2(1 - c_3 s)^2}\right)\psi_n(s) = 0, \tag{B2}$$

satisfying the wave functions

$$\psi_n(s) = \phi(s) y_n(s). \tag{B3}$$

Comparing (B2) with its counterpart (B1), we obtain the following identifications:

$$\tilde{\tau}(s) = c_1 - c_2 s, \quad \sigma(s) = s(1 - c_3 s), \quad \tilde{\sigma}(s) = -\xi_1 s^2 + \xi_2 s - \xi_3, \tag{B4}$$

Following the NU method [18], we obtain the followings [51],

(i) the relevant constant:

$$c_4 = \frac{1}{2}(1 - c_1), \qquad c_5 = \frac{1}{2}(c_2 - 2c_3),$$

$$c_6 = c_5^2 + \xi_1, \qquad c_7 = 2c_4 c_5 - \xi_2,$$

$$c_8 = c_4^2 + \xi_3, \qquad c_9 = c_3(c_7 + c_3 c_8) + c_6,$$

$$c_{10} = c_1 + 2c_4 + 2\sqrt{c_8} - 1 \succ -1, \qquad c_{11} = 1 - c_1 - 2c_4 + \frac{2}{c_3}\sqrt{c_9} \succ -1, \; c_3 \neq 0,$$

$$c_{12} = c_4 + \sqrt{c_8} \succ 0, \qquad c_{13} = -c_4 + \frac{1}{c_3}(\sqrt{c_9} - c_5) \succ 0, \; c_3 \neq 0. \tag{B5}$$

(ii) the essential polynomial functions:

$$\pi(s) = c_4 + c_5 s - \left[\left(\sqrt{c_9} + c_3 \sqrt{c_8}\right)s - \sqrt{c_8}\right], \tag{B6}$$

$$k = -(c_7 + 2c_3 c_8) - 2\sqrt{c_8 c_9}, \tag{B7}$$

$$\tau(s) = c_1 + 2c_4 - (c_2 - 2c_5)s - 2\left[\left(\sqrt{c_9} + c_3 \sqrt{c_8}\right)s - \sqrt{c_8}\right], \tag{B8}$$

$$\tau'(s) = -2c_3 - 2\left(\sqrt{c_9} + c_3 \sqrt{c_8}\right)\langle 0. \tag{B9}$$

(iii) The energy equation:

$$c_2 n - (2n+1)c_5 + (2n+1)\left(\sqrt{c_9} + c_3\sqrt{c_8}\right) + n(n-1)c_3 + c_7 + 2c_3 c_8 + 2\sqrt{c_8 c_9} = 0. \tag{B10}$$

(iv) The wave functions

$$\rho(s) = s^{c_{10}}(1 - c_3 s)^{c_{11}}, \tag{B11}$$

$$\phi(s) = s^{c_{12}}(1 - c_3 s)^{c_{13}}, \; c_{12} > 0, \; c_{13} > 0, \tag{B12}$$

$$y_n(s) = P_n^{(c_{10},c_{11})}(1-2c_3 s), \quad c_{10} > -1, \quad c_{11} > -1, \tag{B13}$$

$$\psi_{n\kappa}(s) = N_{n\kappa} s^{c_{12}} (1-c_3 s)^{c_{13}} P_n^{(c_{10},c_{11})}(1-2c_3 s). \tag{B14}$$

where $P_n^{(\mu,\nu)}(x)$, $\mu > -1$, $\nu > -1$, and $x \in [-1,1]$ are Jacobi polynomials with

$$P_n^{(\alpha,\beta)}(1-2s) = \frac{(\alpha+1)_n}{n!} {}_2F_1(-n, 1+\alpha+\beta+n; \alpha+1; s), \tag{B15}$$

and $N_{n\kappa}$ is a normalization constant. Also, the above wave functions can be expressed in terms of the hypergeometric function as

$$\psi_{n\kappa}(s) = N_{n\kappa} s^{c_{12}} (1-c_3 s)^{c_{13}} {}_2F_1(-n, 1+c_{10}+c_{11}+n; c_{10}+1; c_3 s) \tag{B16}$$

where $c_{12} > 0$, $c_{13} > 0$ and $s \in [0, 1/c_3]$, $c_3 \neq 0$.

**Table 1.** The specific values of the parametric constants for the spin symmetric Dirac-TH problem.

| Constant | Analytic value |
|---|---|
| $c_4$ | $0$ |
| $c_5$ | $-\dfrac{c_h}{2}$ |
| $c_6$ | $\dfrac{c_h^2}{4} + \dfrac{1}{\alpha^2}\left[\kappa(\kappa+1)\left(D_0 c_h^2 - D_1 c_h + D_2\right) + \gamma D r_e^2 + \beta^2 r_e^2 c_h^2\right]$ |
| $c_7$ | $\dfrac{1}{\alpha^2}\left[\kappa(\kappa+1)(-2D_0 c_h + D_1) - 2\gamma D r_e^2 - 2\beta^2 r_e^2 c_h\right]$ |
| $c_8$ | $\dfrac{1}{\alpha^2}\left[-\kappa(\kappa+1)D_0 + \gamma D r_e^2 + 2\beta^2 r_e^2\right]$ |
| $c_9$ | $\dfrac{c_h^2}{4} + \dfrac{1}{\alpha^2}\left[\kappa(\kappa+1)D_2 + \gamma D r_e^2 (c_h-1)^2\right]$ |
| $c_{10}$ | $\dfrac{2}{\alpha}\sqrt{\left[-\kappa(\kappa+1)D_0 + \gamma D r_e^2 + 2\beta^2 r_e^2\right]}$ |
| $c_{11}$ | $\dfrac{2}{c_h}\sqrt{\dfrac{c_h^2}{4} + \dfrac{1}{\alpha^2}\left[\kappa(\kappa+1)D_2 + \gamma D r_e^2 (c_h-1)^2\right]}$ |
| $c_{12}$ | $\dfrac{1}{\alpha}\sqrt{\left[-\kappa(\kappa+1)D_0 + \gamma D r_e^2 + 2\beta^2 r_e^2\right]}$ |
| $c_{13}$ | $\dfrac{1}{2} + \dfrac{1}{c_h}\sqrt{\dfrac{c_h^2}{4} + \dfrac{1}{\alpha^2}\left[\kappa(\kappa+1)D_2 + \gamma D r_e^2 (c_h-1)^2\right]}$ |

**Table 2** The Dirac energy spectrum, $E_{n,\kappa}$ (in $fm^{-1}$), for various quantum numbers $n$ and $\kappa$ in the presence of spin symmetry.

| $l$ | $n, \kappa<0, \kappa>0$ | $(l, j = l+1/2)$ | $c_h = 0^{(II)}$ Present | $c_h = 0^{(II)}$ [44] | $c_h = 0^{(I)}$ Present | $c_h = 0.01$ Present |
|---|---|---|---|---|---|---|
| 1 | 0, -2,1 | $0p_{3/2}, 0p_{1/2}$ | 0.0188481 | 0.0188481 | 0.0158972 | 0.0156445 |
| 2 | 0, -3,2 | $0d_{5/2}, 0d_{3/2}$ | 0.0336562 | 0.0336562 | 0.0289087 | 0.0292850 |
| 3 | 0, -4,3 | $0f_{7/2}, 0f_{5/2}$ | 0.0525273 | 0.0525273 | 0.0454736 | 0.0468568 |
| 4 | 0, -5,4 | $0g_{9/2}, 0g_{7/2}$ | 0.0754350 | 0.0754350 | 0.0655857 | 0.0683657 |
| 1 | 1, -2,1 | $1p_{3/2}, 1p_{1/2}$ | 0.0899995 | 0.0899995 | 0.0721426 | 0.0711732 |
| 2 | 1, -3,2 | $1d_{5/2}, 1d_{3/2}$ | 0.1136725 | 0.1136725 | 0.0933683 | 0.0926634 |
| 3 | 1, -4,3 | $1f_{7/2}, 1f_{5/2}$ | 0.1438031 | 0.1438031 | 0.120011 | 0.119939 |
| *4* | 1, -5,4 | $1g_{9/2}, 1g_{7/2}$ | 0.1791425 | 0.1791425 | 0.151061 | 0.152013 |

**Table 3** The Dirac hole energy states, $E_{n,\kappa}$ (in $fm^{-1}$), for various quantum numbers $n$ and $\kappa$ in the presence of pspin symmetry.

| $\tilde{l}$ | $n, \kappa<0, \kappa>0$ | $(l, j)$ | $c_h = 0^{(II)}$ Present | $c_h = 0^{(II)}$ [43] | $c_h = 0^{(I)}$ Present | $c_h = -0.01$ Present |
|---|---|---|---|---|---|---|
| 1 | 1, -1,2 | $1s_{1/2}, 0d_{3/2}$ | −0.0064123 | −0.0064123 | −0.0063644 | −0.0078235 |
| 2 | 1, -2,3 | $1p_{3/2}, 0f_{5/2}$ | −0.0155771 | −0.0155771 | −0.0152135 | −0.0192390 |
| 3 | 1, -3,4 | $1d_{5/2}, 0g_{7/2}$ | −0.0243659 | −0.0243659 | −0.0233169 | −0.0308043 |
| 4 | 1, -4,5 | $1f_{7/2}, 0h_{9/2}$ | −0.0305297 | −0.0305297 | −0.0285678 | −0.0403430 |
| 1 | 2, -1,2 | $2s_{1/2}, 1d_{3/2}$ | −0.0070204 | −0.0070204 | −0.0070051 | −0.0085285 |
| 2 | 2, -2,3 | $2p_{3/2}, 1f_{5/2}$ | −0.0190441 | −0.0190441 | −0.0188890 | −0.0232805 |
| 3 | 2, -3,4 | $2d_{5/2}, 1g_{7/2}$ | −0.0337719 | −0.0337719 | −0.0331986 | −0.0415466 |
| *4* | 2, -4,5 | $2f_{7/2}, 1h_{9/2}$ | −0.0492150 | −0.0492150 | −0.0478538 | −0.0611045 |

**Table 4.** Model parameters of the diatomic molecules studied in the present work

| Molecule | $c_h$ | $\mu$ (amu) | $b_h(A^{\circ-1})$ | $r_e(A^\circ)$ | $D(cm^{-1})$ |
|---|---|---|---|---|---|
| $H_2$ [12] | 0.170066 | 0.50391 | 1.61890 | 0.741 | 38318 |
| $I_2$ [12] | -0.139013 | 10.612 | 2.12343 | 2.666 | 12547 |

**Table 5.** Relativistic energy eigenvalues (in eV) for $H_2$ and $I_2$ diatomic molecules using the spin symmetry TH potential for various values of $n$, $\kappa$ and $c_s = \mu$.

| $l$ | $n, \kappa$ | $(l, j = l+1/2)$ | $H_2$ | $I_2$ |
|---|---|---|---|---|
| 0 | 1,-1 | $1s_{1/2}$ | 4.496299243 | 0.04301938173 |
| 1 | 1,-2 | $1p_{3/2}$ | 4.792825206 | 0.04908477248 |
| 2 | 1,-3 | $1d_{5/2}$ | 5.265998324 | 0.06198365883 |
| 0 | 2,-1 | $2s_{1/2}$ | 5.208297483 | 0.1330310673 |
| 1 | 2,-2 | $2p_{3/2}$ | 5.393734566 | 0.1391726596 |
| 2 | 2,-3 | $2d_{5/2}$ | 5.714484641 | 0.1517415539 |

**Table 6.** Relativistic energy eigenvalues (in eV) for $H_2$ and $I_2$ diatomic molecules using the pspin symmetry TH potential for various values of $n$, $\kappa$ and $c_{ps} = -\mu$.

| $l$ | $n, \kappa$ | $(l, j = l+1/2)$ | $H_2$ | $I_2$ |
|---|---|---|---|---|
| 0 | 1,-1 | $1s_{1/2}$ | −4.716308462 | −0.04908477248 |
| 1 | 1,-2 | $1p_{3/2}$ | −5.219487600 | −0.06198365885 |
| 2 | 1,-3 | $1d_{5/2}$ | −5.808371132 | −0.0828077168 |
| 0 | 2,-1 | $2s_{1/2}$ | −5.377765079 | −0.1391726596 |
| 1 | 2,-2 | $2p_{3/2}$ | −5.738903598 | −0.1517415539 |
| 2 | 2,-3 | $2d_{5/2}$ | −6.198359366 | −0.1712385573 |

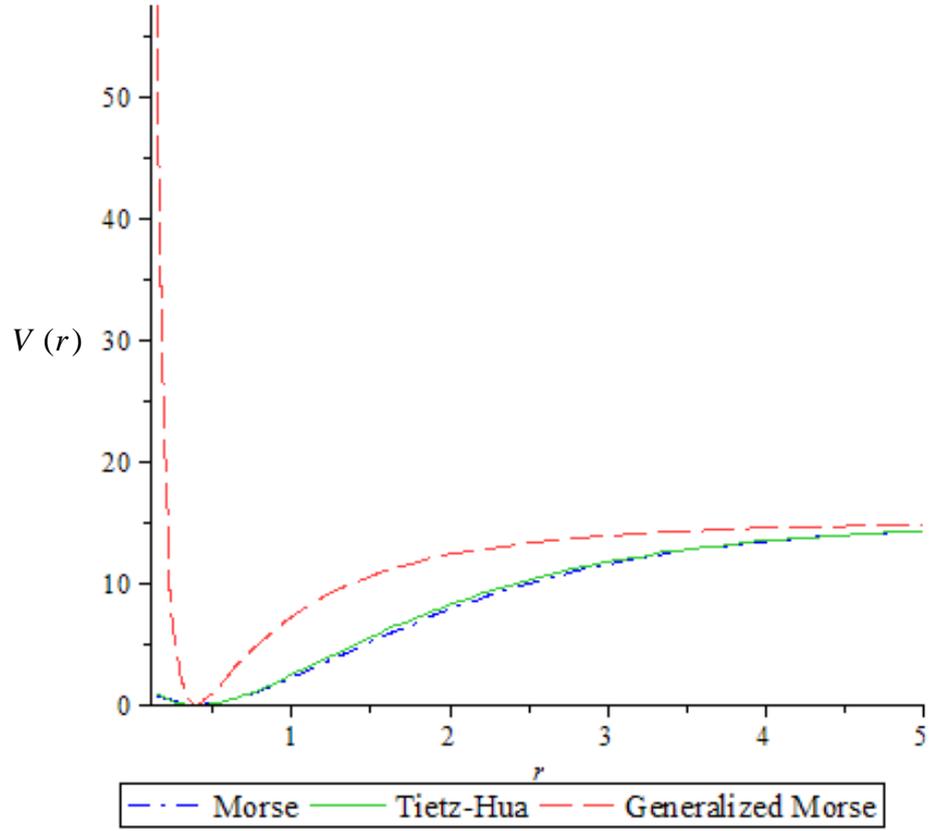

**Fig. 1.** The shape of potentials discussed in this work.

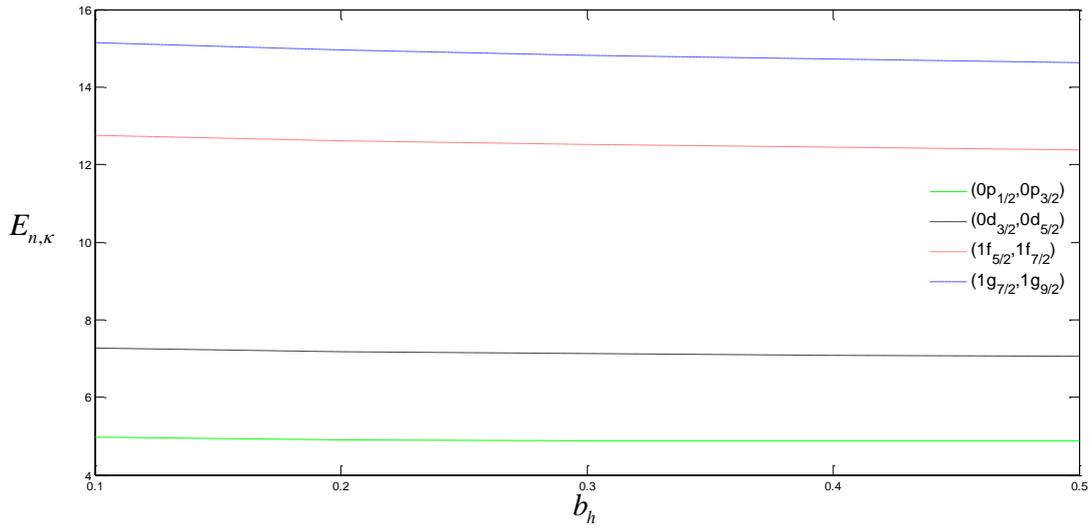

Fig. 2. The variation of the energy levels as a function $b_h$ in the presence of spin symmetry.

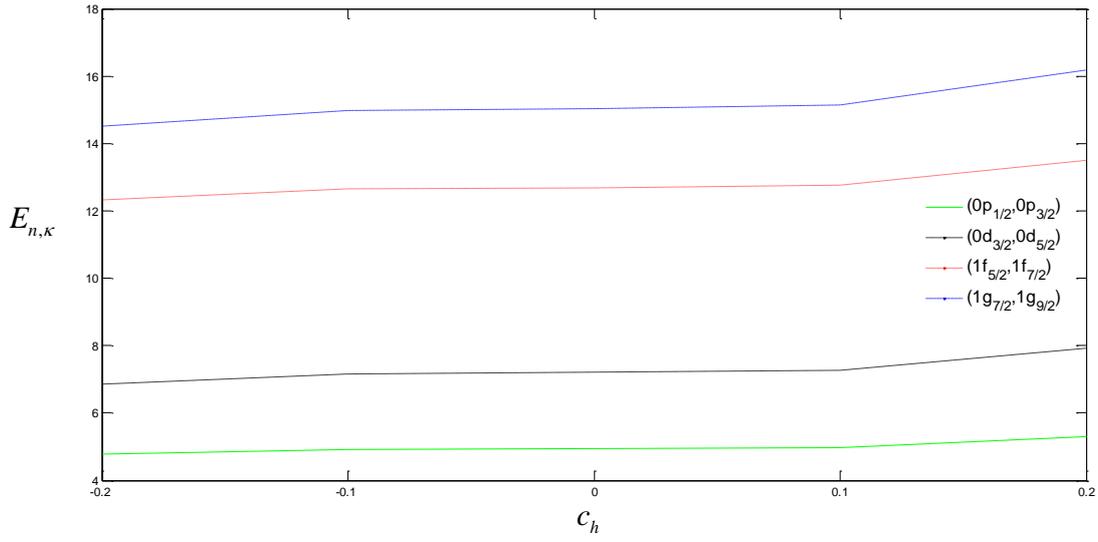

Fig. 3. The variation of the energy levels as a function $c_h$ in the presence of spin symmetry.

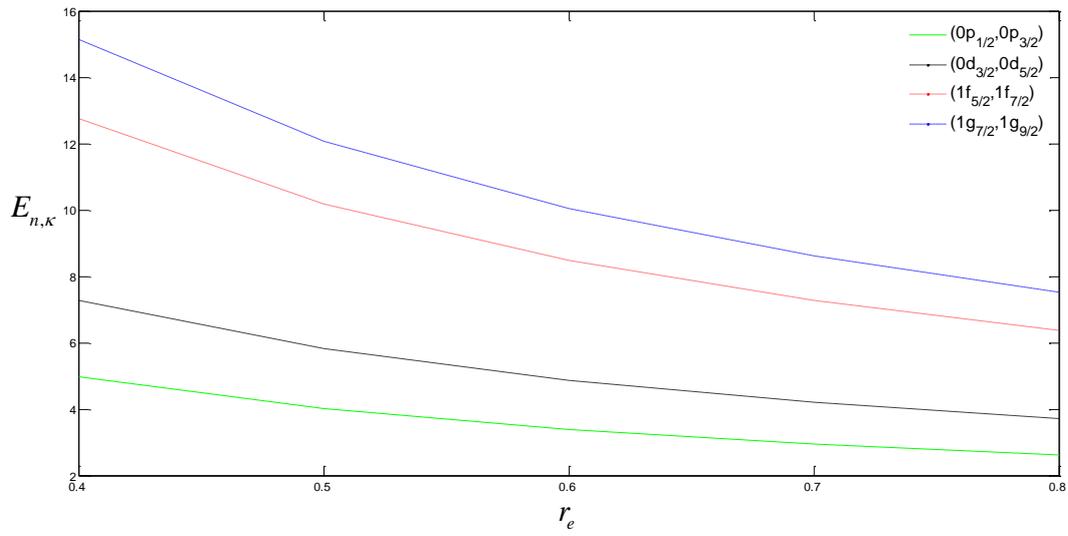

Fig. 4. The variation of the energy levels as a function $r_e$ in the presence of spin symmetry.

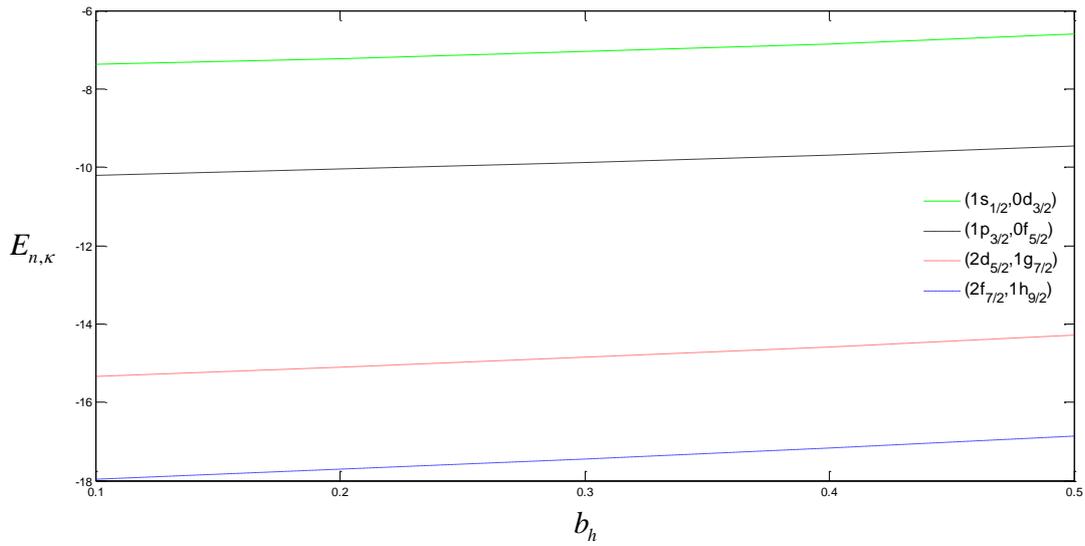

Fig. 5. The variation of the Dirac hole energy states as a function $b_h$ in the presence of pspin symmetry.

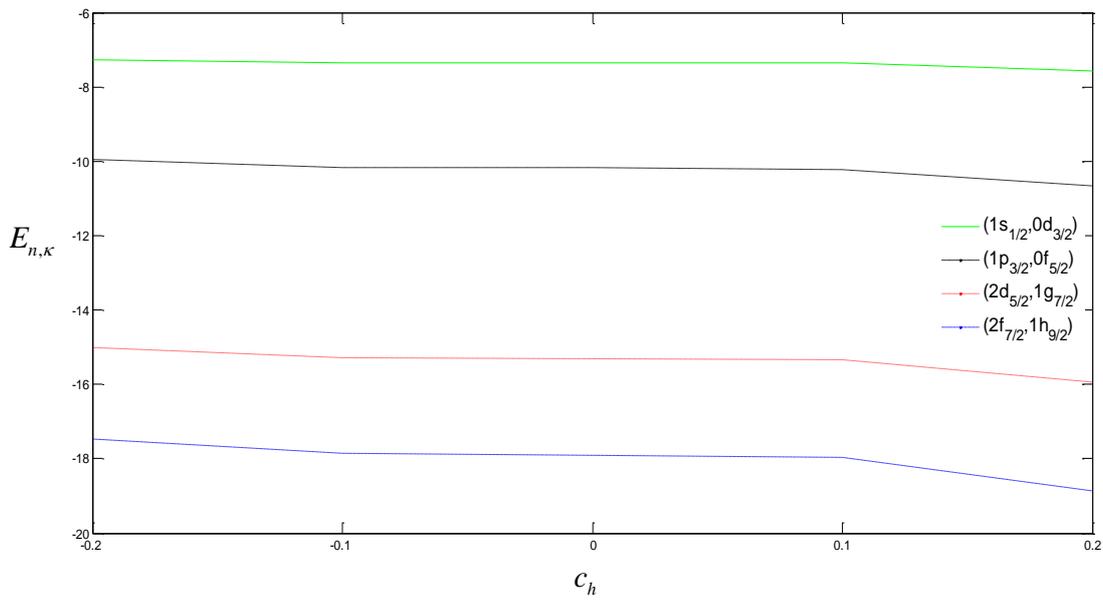

Fig. 6. The variation of the Dirac hole energy states as a function $c_h$ in the presence of pspin symmetry.

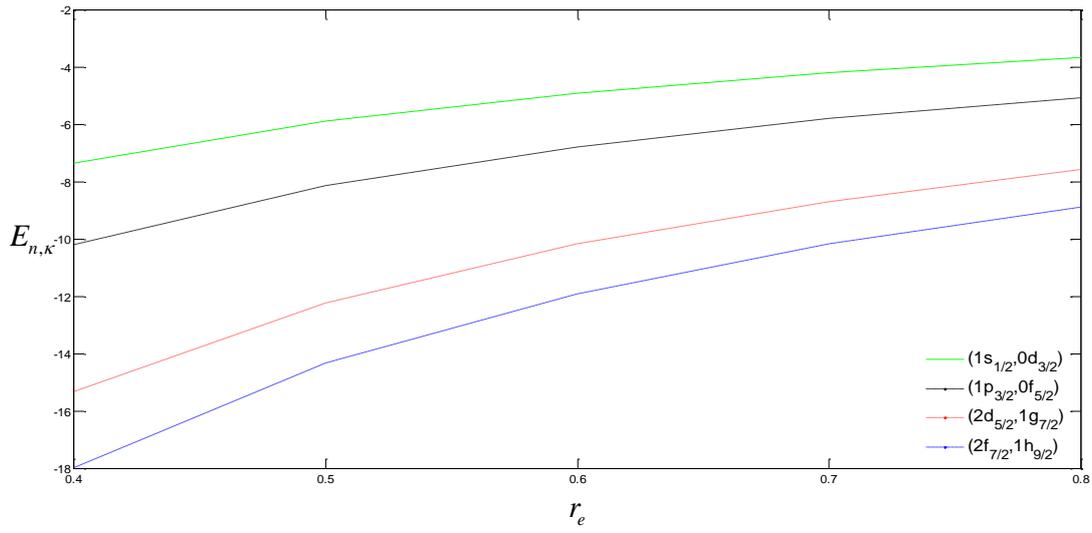

Fig. 7. The variation of the Dirac hole energy states as a function $r_e$ in the presence of pspin symmetry.